\documentclass[preprints,article,accept,pdftex,moreauthors]{Definitions/mdpi} 

\firstpage{1} 
\pubvolume{1}
\issuenum{1}
\articlenumber{0}
\pubyear{2023}
\copyrightyear{2023}
\datereceived{ } 
\daterevised{ } 
\dateaccepted{ } 
\datepublished{ } 
\hreflink{https://doi.org/} 

\usepackage[utf8]{inputenc}

\usepackage[normalem]{ulem}

\usepackage{notoccite}

\newcommand{\exv}[1]{\left\langle #1 \right\rangle}
\newcommand{\pwisein}{\left\{ \begin{array}{ll}}
\newcommand{\pwiseout}{\end{array}\right.}

\Title{Size-pore-dependent methanol sequestration from water-methanol mixtures by an embedded graphene slit}

\TitleCitation{Size-pore-dependent methanol sequestration from water-methanol mixtures by an embedded graphene slit}

\Author{Roger Bellido-Peralta$^{1,\dagger}$\orcidA{}, 
Fabio Leoni$^{2}$\orcidB{},  
Carles Calero$^{1,3}$\orcidC{},
and
Giancarlo Franzese$^{1,3}\orcidD{}$ 
}

\AuthorNames{Roger Bellido-Peralta, Fabio Leoni and Carles Calero, Giancarlo Franzese}

\AuthorCitation{Bellido-Peralta, R.; Leoni, F; Calero, C.; Franzese, G.}

\address{%
$^{1}$ \quad Secci\'o de F\'isica Estad\'istica i Interdisciplin\`aria - Departament de F\'isica de la Mat\`eria Condensada, Universitat de Barcelona, Mart\'{\i} i Franqu\`es 1, 08028 Barcelona, Spain\\
$^{2}$ \quad Department of Physics, Sapienza University of Rome, Italy; \\
$^{3}$ \quad Institut de Nanoci\`{e}ncia i Nanotecnologia, Universitat de Barcelona, 08028 Barcelona, Spain.
}

\corres{Correspondence: gfranzese@ub.edu;
}

\firstnote{Currently at National Graphene Institute, University of Manchester, UK; roger.bellido@postgrad.manchester.ac.uk} 

\abstract{
The separation of liquid mixture components is relevant in many applications--going from water purification to biofuel
production--and a growing concern related to the UN Sustainable Development Goals (SDGs), such as ``Clean water and Sanitation'' and ``Affordable and clean energy''.
One promising technique is using graphene slit-pores as filters, or sponges, because the confinement potentially affects the properties of the mixture components in different ways, favoring their separation. 
However, no systematic study shows how the size of a pore changes the thermodynamics of the surrounding mixture.
Here, we focus on water-methanol mixtures and explore, using Molecular Dynamics simulations, the effects of a graphene pore, with size ranging from 6.5 to 13 \AA,
for three compositions: pure water, 90\%-10\%, and 75\%-25\% water-methanol. 
We show that tuning the pore size can change the mixture pressure, density, and composition in bulk due to the size-dependent methanol sequestration within the pore.
Our results can help in optimizing the graphene pore size for filtering applications.  
}


\begin{document}

\section{Introduction}

According to the latest UN progress report on accomplishing the Sustainable Development Goals (SDGs), millions still need 'clean water and sanitation' (SDG 6) and 'affordable and clean energy' (SDG 7). As a result, meeting the targets by the 2030 deadline is currently unreachable. Therefore, there is an urgent need for more research and investment from governments and businesses to accelerate the implementation of these goals and ensure a sustainable future for everyone. In particular, SDG 6 and SDG 7 are particularly relevant for addressing some of the most pressing challenges of our time. SDG 7 seeks to ensure access to affordable, reliable, sustainable, and modern energy for all. This is essential for reducing greenhouse gas emissions \cite{Fetisov_2023}, such as the byproducts of the oil industry \cite{Fetisov2023}, improving health and well-being, enhancing economic productivity, and creating opportunities for innovation and social inclusion. SDG 6 aims to ensure the availability and sustainable management of water and sanitation for all. This is vital for preventing diseases, improving hygiene, reducing inequalities, protecting ecosystems, and supporting human dignity. Furthermore, they are directly related to SDG 3 (Good health and well-being), SDG 10 (Reduced inequalities), SDG 11 (Sustainable cities and communities), SDG 12 (Responsible consumption and production), SDG 13 (Climate action), SDG 14 (Life below water), SDG 15 (Life on land), and indirectly to all the others.

Finding new and efficient ways to separate water from methanol is relevant in this context. Indeed, water and methanol are often mixed in various industrial processes, such as biodiesel production, wastewater treatment, and solvent extraction. However, separating water and methanol is challenging and energy-intensive, as they form an azeotropic mixture that cannot be easily distilled. By developing more efficient and cost-effective separation methods, such as membrane technology, adsorption, or extraction, these processes' energy consumption and environmental impact can be reduced, contributing to the goal of affordable and clean energy. Furthermore, by recovering water and methanol from these mixtures, the quality and quantity of water resources can be improved, as well as the availability of methanol as a renewable fuel or chemical feedstock, contributing to the goal of clean water and sanitation.

Water and methanol are fully miscible liquids at ambient conditions due to their hydrogen bonds \cite{Soetens:2015aa}, and their mixtures are standard in food processing, preservation, pharmaceutical, and chemical industries. For example, water-methanol blends are used in power generation applications, including gas turbines, fuel cells, green alternative fuels, improved combustion engines, and solar plants \cite{Ren_2000, Boysen_2004, Liu_2007, Yanju_2008, Miganakallu:2020aa, Toledo-Camacho:2021aa}. Also, methanol is often added to water to lower its freezing point and improve its flow \cite{Miller:1964aa, Sun:2011aa}. Nevertheless, it is necessary to separate the two components in several applications. For example, separating water and methanol is essential in producing biofuels to ensure the quality and efficiency of the final product \cite{Cortright:2002aa, Masoumi:2021aa}, or in chemical manufacturing processes to maintain the desired concentration of the reactants and prevent unwanted side reactions \cite{Dalena:2018aa}. 

However, the separation of methanol from water is usually performed by inefficient and energetically-intensive distillation \cite{Liang_2014}. Therefore, to minimize energy waste and increase efficiency, researchers have investigated water purification via nanomembrane filtering using chemical functionalization \cite{Azamat:2019ua, Azamat:2021aa}, adsorption on graphite pores \cite{Prslja:2019aa}, or infinite graphite sheets \cite{Mosaddeghi:2019aa}. In general, nanomaterials and nanomembranes combined with advanced catalytic, photothermal, adsorption, and filtration processes provide fast, efficient, and tunable alternatives compared to conventional routes in water remediation. However, many challenges regarding scalability and sustainability are still open \cite{Esplandiu:2023aa}.

Recently, graphene-based membranes have been reported \cite{Nair2012} as a method that changes the dynamics of the confined fluids when compared to bulk. Furthermore, several works have been published exploring all types of materials, looking for the selectivity of water or methanol over the other component of the mixture \cite{Mahmood2012, Villegas2015, Hung2017, Kachhadiya2021}. Also, Molecular Dynamics (MD) simulations of atomistic models clarified the physical mechanisms of these changes \cite{calero2020} and the differences among various typologies of fluids \cite{Leoni:2021aa, LF2014, Leoni:2016aa}. Table \ref{tab:results} summarizes some of the recently obtained main experimental and theoretical results.

\begin{table}
\begin{tabular}{| m{2.2cm} | m{1cm} | m{1.8cm} | m{6cm} | m{0.5cm} |}
\hline
Material & Method & Selective to & Main Result & Ref. \\ \hline \hline
\begin{tabular}[c]{@{}l@{}}Porous BNNS \\ -H, -F \\ -OH \end{tabular}& \begin{tabular}[c]{@{}l@{}} Simu-\\ lations \end{tabular}& \begin{tabular}[c]{@{}l@{}} Methanol \\ Water \end{tabular}& Each molecule has higher free energy in correspondence to the pores it cannot permeate through. & \cite{Azamat:2019ua} \\ \hline
BNNT & \begin{tabular}[c]{@{}l@{}} Simu-\\ lations \end{tabular}& Alcohols & Alcohols can easily break their hydrogen bonds to enter and occupy the nanotubes,  having a strong interaction with them. & \cite{Azamat:2021aa} \\ \hline
Pristine graphene & \begin{tabular}[c]{@{}l@{}} Simu-\\ lations \end{tabular}& Methanol & Methanol gets preferentially absorbed into a graphene slit pore. When mixed with water, the two liquids couple and diffuse. & \cite{Prslja:2019aa} \\ \hline
Graphite plates & \begin{tabular}[c]{@{}l@{}} Simu-\\ lations \end{tabular}& Methanol & Preferential absorption of methanol on graphite sheets due to Van der Walls interactions between the methyl groups and the carbon.  & \cite{Mosaddeghi:2019aa} \\ \hline

GO & \begin{tabular}[c]{@{}l@{}} Experi-\\ ments \end{tabular}& Water & Low friction flow of a water monolayer through 2D channels between graphene sheets, while helium remains in feed. & \cite{Nair2012} \\ \hline
SA/PVA  & \begin{tabular}[c]{@{}l@{}} Experi-\\ ments \end{tabular}& Water        & $\uparrow$ T $\Rightarrow$ $\uparrow$ mobility of the polymer chain$\Rightarrow$ $\uparrow$ flux, little selectivity reduction. At 5\% PVA composition, the material has surface pores, and at 20\% has cracks. Optimum PVA composition at 10\%. &  \cite{Mahmood2012}    \\ \hline
PHB  & \begin{tabular}[c]{@{}l@{}} Experi-\\ ments \end{tabular}& Water        & Pure substance pervaporation shows good MeOH permeation. It has water selectivity in a mixture due to the hydrogen bond network. MeOH has reduced mobility when mixed with water. & \cite{Villegas2015}     \\ \hline
rGO/CS  & \begin{tabular}[c]{@{}l@{}} Experi-\\ ments \end{tabular}& Water        & The interlayer space due to the CS leads to molecularly sieve water, and the hydrophobicity of GO provides good flux. & \cite{Hung2017}     \\ \hline
\begin{tabular}[c]{@{}l@{}}ZIF-8/PVDF \\ ZIF-67/PVDF\end{tabular} & \begin{tabular}[c]{@{}l@{}} Experi-\\ ments \end{tabular}& Water        & ZIF-67/PVDF membrane enhances flux due to its hydrophilicity. However, $\uparrow$ water \% in the feed $\Rightarrow$ $\uparrow$ swelling $\Rightarrow$ $\downarrow$ selectivity as volume increases and MeOH molecules can also pass through. $\uparrow$ T $\Rightarrow$ $\uparrow$ polymer chain mobility $\Rightarrow$ $\uparrow$ flux, $\downarrow$ selectivity. & \cite{Kachhadiya2021}     \\ \hline

\end{tabular}
\caption{\textbf{Summary of experimental or simulation results with different membranes for water-methanol mixtures.} List of acronyms: SA (sodium alginate), PVA (polyvinyl alcohol), PHB (poly(3-hydroxybutyrate)), rGO (reduced Graphene Oxide), CS (chitosan), ZIF-n (Zeolitic Imidazolate Framework, where the number "n" is not related to the structure, just used as naming \cite{Park2006}), PVDF (polyvinylidene fluoride), BNNS (boron nitride nanosheets, functionalized with -H,-F groups and -OH group), BNNT (boron nitride nanotubes).
Symbols $\uparrow$, $\downarrow$, $\Rightarrow$ stands for {\it increasinging, decreasing, implying}, respectively.}
\label{tab:results}
\end{table}

In particular, water has peculiar properties that are anomalous compared to other fluids  \cite{life, fermi2012, Gallo:2021wx}, and its interaction with nanointerfaces dramatically modifies its structure \cite{MCF2017}, thermodynamics and dynamics \cite{calero2020}, leading to unusual transport properties for both water and solutes \cite{Corti:2021uy}.

On the other hand, methanol, the smallest alcohol, has an apolar methyl group (CH$_3$) and a polar hydroxyl group (OH). The polar moiety can form hydrogen bonds of strength and length similar to water which, together with the methanol's small size, allows it to fully integrate into the water's hydrogen bond network \cite{Hus:2014aa}. 

Previous works have shown that, under slit-pore confinement, water's thermal diffusion coefficient $D_{\|}$ parallel to the walls is non-monotonic when the pore width $\delta$ changes below 1.5 nm \cite{calero2020}. 
This property is a consequence of water's ability to form hydrogen bonds. However, recent atomistic simulations show that this behavior is also present in liquids without hydrogen bonds, including simple van der Waals liquids or not-network-forming anomalous liquids \cite{Leoni:2021aa}. Nevertheless, the study shows that the mechanism leading to the variation of $D_{\|}$ in confined water is unique and different from other liquids  \cite{Leoni:2021aa}. Therefore, using nano-confining graphene slit-pores to separate water from methanol based on physical processes is an appealing possibility.

However, no systematic study shows how the size of a graphene slit-pore changes the composition and thermodynamics of the surrounding mixture in which it is embedded as a solute and how to optimize it for filtering applications. 
This geometry reminds the recent application of nanoengineered graphene pores used as a {\it sponge} for overcoming the limitations of the existing water treatment systems \cite{Esplandiu:2023aa,
Baptista_Pires_2021}.

Here, we investigate the capacity of a graphene slit-pore to sequester methanol from a water-methanol mixture and its effects on the mixture. In particular, we study how different water-methanol mixtures' properties-such as density, pressure, and composition- are affected as the width $\delta$ of an embedded graphene slit-pore changes. 
To this end, we perform Molecular Dynamics (MD) simulations of water-methanol mixtures of different compositions with graphene slit-pores of different sizes as solutes. 
The water-methanol mixture is described using tested coarse-grained models (see Sec.~\ref{sec:Methodology}) based on Continuous Shouldered Well (CSW) and Lennard-Jones potentials. 

\section{Results and Discussion}

\subsection{Number density}

\subsubsection{The pure CSW case.}

\begin{figure}
\centering
\includegraphics[width=1\columnwidth]{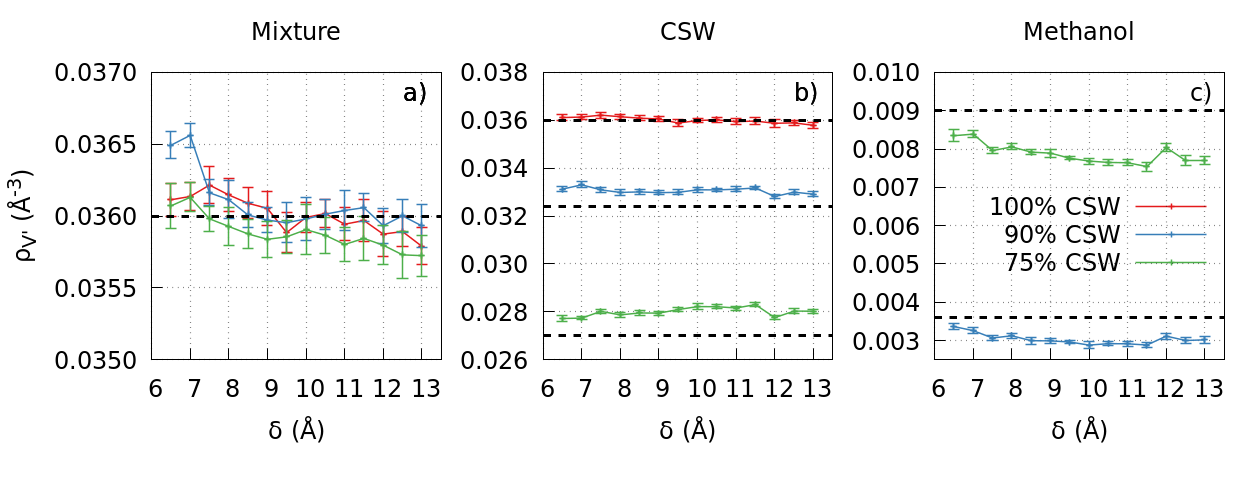}
\caption{{\bf  The slit-pore width affects the densities of the mixture outside the pore.}
For pure CSW (red) and CSW-methanol compositions 90\%-10\% (blue) and 75\%-25\% (green), the change in $\delta$ implies variations of density $\rho_{V'}$ of the mixture (panel a) and each component (CSW, panel b, and methanol, panel c) outside the pore. 
In each panel, horizontal dashed lines mark the values of $\rho_{V'}$ that would be expected without the embedded slit-pore for the mixtures with different compositions as indicated in the legend in panel c, for the mixture (a), the CSW (b), the methanol (c).
}
\label{figbulk}
\end{figure}

First, we check the behavior of the pure CSW in the subregion $V'$ outside the pore (Fig. \ref{figbulk}a) and find that the slit-pore width $\delta$ affects weakly its density (Fig.~\ref{figbulk}a, b). In particular, the CSW density in $V'$ has a minor decrease 
for increasing $\delta$, but remains close to the overall density of 0.036 $\textup{\r{A}}^{-3}$ within the error, which is the nominal number density in the entire simulation box, including the slit-pore (see Sec.~\ref{sec:Methodology}).
Therefore, for the considered range of $\delta$, the nano-pore does not adsorb much water-like liquid inside, consistent with its hydrophobic properties at the macroscopic scale.

Nevertheless, in Ref.~\cite{Leoni:2021aa} a similar confined liquid (CSW with $\Delta=30$) shows free energy and confined density extrema at 
$\sim7.5\ \textup{\r{A}}$,
$\sim9.5\ \textup{\r{A}}$,
$\sim10.5\ \textup{\r{A}}$,  
and 
$\sim12\  \textup{\r{A}}$ that seems to correlate with the small density variations we find here, although within the error bars. Hence, we explore, next, the behavior of the mixture in $V'$ to better investigate the pore-size effects.

\subsubsection{The mixture case.}

At some fixed pore sizes, e.g., $\delta=9$ \AA, we find an overall decrease for the mixture density $\rho_{V'}$ when we increase the methanol concentration from 0\% to 10\% and 25\% (Fig.~\ref{figbulk} a). 
This behavior is consistent with what is expected for bulk. \cite{Sentenac:1998aa, perry1999perry}. At other values of $\delta$, e.g., $\delta=7$ \AA, the trend is nonmonotonic in methanol concentration, suggesting an unexpected behavior.

\begin{figure}
\centering
\includegraphics[width=1\columnwidth]{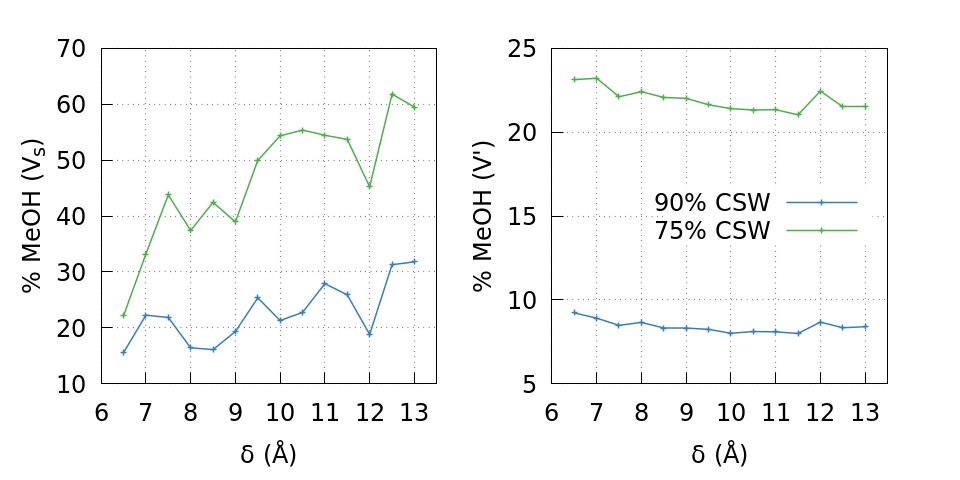}
\caption{{\bf  The sequestration of methanol inside the graphene slit-pore affects its concentration outside.}
The methanol concentrations inside (left panel) and outside the pore (right panel) are anti-correlated for 
both mixture compositions 90\%-10\% (blue) and 75\%-25\% (green) CSW-methanol. 
Inside the pore, the concentration can increase by $\simeq$ 320\%, in the first case, compared to its nominal bulk value and 250\%, in the second case, while outside can be $\simeq$ 75\% or 84\% lower, respectively.
}
\label{figmet}
\end{figure}

Indeed, surprisingly, we observe that changes in the slit pore width $\delta$ affect the mixture $\rho_{V'}$ outside the error bars. Moreover, the effect in the mixture is more evident than in pure CSW liquid. Therefore, the pore-size dependence of the density in $V'$ depends mainly on the methanol-pore interaction.

To better understand this dependence, we calculate the concentration of each component in $V'$ separately as a function of $\delta$.
The overall CSW number density in the subvolume $V'$, $\rho_{CSW}$, should decrease from 0.036 $\textup{\r{A}}^{-3}$ at 0\% methanol concentration, to 0.0324 $\textup{\r{A}}^{-3}$ at 10\% methanol, to 0.027 $\textup{\r{A}}^{-3}$ at 25\% methanol.
However, we find that at the highest methanol concentration, $\rho_{CSW}$ is $\approx 4\%$ above the expected value (Fig.~\ref{figbulk} b). 
Therefore, there is less methanol in the solution than expected due to the presence of the graphene pore.

Furthermore, $\rho_{CSW}$, at nominal 25\% methanol concentration, tends to increase for larger pore sizes $\delta$ (Fig.~\ref{figbulk} b). Hence, the amount of methanol in the mixture in $V'$ decreases for greater $\delta$, and the strength of the effect is proportional to the nominal methanol concentration.

This conclusion is confirmed when we explicitly calculate the methanol number density in $V'$, $\rho_{meth}$, (Fig.~\ref{figbulk} c). The expected values for $\rho_{meth}$ 
are 0.0036 $\textup{\r{A}}^{-3}$ at 10\% methanol concentration and 0.0090 $\textup{\r{A}}^{-3}$ at 25\% methanol. However, we find that $\rho_{meth}$ can be as low as $\approx 83\%$ of the expected values.

In particular, the decrease of $\rho_{meth}$ in both cases is non-monotonic, with maxima and minima that partially correlate with the confined CSW density extrema at 
$\sim7.5\ \textup{\r{A}}$,
$\sim9.5\ \textup{\r{A}}$,
$\sim10.5\ \textup{\r{A}}$,  
and 
$\sim12\  \textup{\r{A}}$
found in Ref.~\cite{Leoni:2021aa}, suggesting that the polar part of the methanol interaction could partially be responsible for this non-monotonicity.
Furthermore, the decrease is more evident when the nominal methanol concentration is higher, in agreement with the $\rho_{CSW}$ behavior.
Finally, the overall $\rho_{meth}$ decreases for increasing $\delta$, suggesting an increased absorption of methanol inside the graphene pore when its size increases.

To test this result directly, we calculate the methanol sequestered by the pore inside walls as a function of $\delta$ for the two mixture compositions (Fig.~\ref{figmet}, left panel).
We find that methanol in the pore exceeds what would be expected from simple osmotic equilibrium. 
In particular, at the overall 10\% and 25\% methanol compositions, we find up to 320\% and 250\% more methanol than expected inside the pore, respectively.
Furthermore, the greater $\delta$, the higher the methanol sequestered, above the value expected by simple osmotic diffusion. Hence, the pore inside adsorbs more methanol when its size increases.

Consequently, the
methanol amounts inside and outside the pore are anti-correlated. The effect in $V'$ is more substantial for higher methanol overall concentration (Fig.~\ref{figmet}). 
For increasing slit size, the trend for the methanol concentration is to increase inside and decrease outside the pore. However, the methanol concentration changes non-monotonically, correlating with the oscillatory properties of the confined CSW liquid \cite{Leoni:2021aa}.

\subsection{Pressure in $V'$}

\begin{figure}
\centering
\includegraphics[width=0.7\columnwidth]{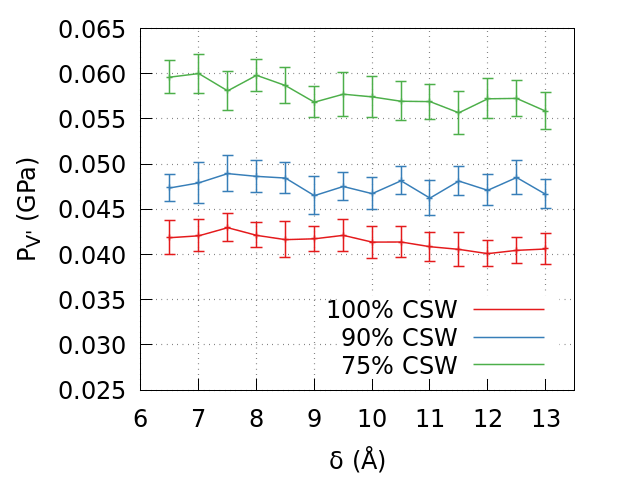}
\caption{{\bf  The pressure in $V'$ at different mixture compositions and graphene slit-pore sizes.} $P_{V'}$ increases when the methanol concentration goes from 0\% to 10\%, to 25\%, and has a weak dependence on the pore's width $\delta$. The dependence is more evident for higher methanol concentrations.
}
\label{figp_bulk}
\end{figure}

Our results show a possible correlation between the observed variation of the mixture density in $V'$ and the properties of the confined CSW model adopted for the polar interactions of the mixture components. Although the CSW model has been tested in the literature for the mixture bulk-properties \cite{Marques:2020aa}, there is no study about its 
reliability when the bulk embeds a graphene pore. 
Therefore, we calculate the mixture's pressure $P_{V'}$ in the subvolume $V'$ to test if the coarse-grained model qualitatively reproduces the correct thermodynamics (Fig.\ref{figp_bulk}).

First, we find that $P_{V'}$ increases as we increase the amount of methanol in the system. This is consistent with the expected thermodynamics  \cite{Sentenac:1998aa, perry1999perry}.

At 0\% and low, 10\%, methanol concentration, 
the overall $P_{V'}$ does not change significantly within the error bars when the slit-pore width varies. 
However,  for 25\% methanol concentration, the $P_{V'}$ tends to decrease weakly for increasing pore's width $\delta$. 

\begin{figure}
\centering
\includegraphics[width=0.7\columnwidth]{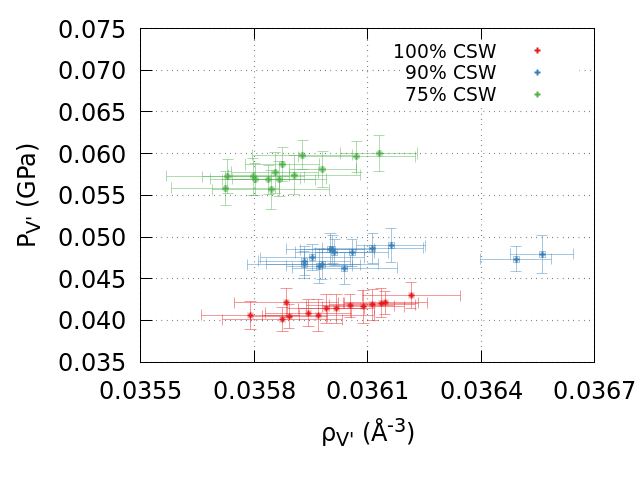}
\caption{{\bf The pressure and density of the mixture in $V'$ behave as expected.}
We calculate the density $\rho_{V'}$ and pressure $P_{V'}$ in $V'$ for pure CSW (red) and CSW-methanol mixtures with compositions 90\%-10\% (blue) and 75\%-25\% (green) for different slit-pore widths (Fig.~\ref{figbulk}a and Fig.~\ref{figp_bulk}, respectively). The parametric plot shows that 
the two quantities are proportional and, at fixed $\rho_{V'}$, $P_{V'}$ increases for increasing methanol concentration, consistent with experiments in bulk \cite{Sentenac:1998aa, perry1999perry}.
}
\label{figp_vs_d}
\end{figure}

Because our calculations show that the density of the mixture in $V'$ weakly oscillates with $\delta$ (Fig.~\ref{figbulk} a), to test the mixture's equation of state, at least qualitatively, we make a parametric plot of $P_{V'}$ as a function of $\rho_{V'}$ (Fig.~\ref{figp_vs_d}).
Within the range of the observed $\delta$-dependent variations, 
we find a behavior that is qualitatively consistent with the existing experimental results, being $P_{V'}$ an increasing function of $\rho_{V'}$ that is approximately linear within the error bars
 \cite{Sentenac:1998aa, perry1999perry}.

Therefore, the coarse-grained model preserves the mixture equation of state qualitatively under the thermodynamic conditions we explore here. This occurs despite the lack of directionality in the polar interaction. As discussed in Ref.~\cite{Vilaseca2011}, a consequence of this approximation
 is the incorrect estimate of the polar entropic contribution to the free energy of the mixture, as demonstrated in Ref.~\cite{Leoni:2021aa} when comparing the free energy of the confined CSW potential with that of the atomistic TIP4P/2005 water model.

\section{Materials and Methods}
\label{sec:Methodology}

In this study, we use a coarse-grained model for the water-methanol mixture \cite{Marques:2020aa} in which both molecules are represented schematically as beads, one for water and two for methanol. Here, the OH is modeled using the Continuous Shouldered Well (CSW) potential \cite{Fr07a}, which has been used to reproduce several properties of systems as different as liquid metals, colloids, or, as in our case of interest, water (and hydroxyl functional groups) \cite{Hus:2014aa, Hus2014}.

Although it does not reproduce all the water properties due to its lack of directionality \cite{Vilaseca2011}, the CSW represents a simple approximation that significantly reduces the computation time in MD simulations compared to atomistic models. Furthermore, it is simple enough to be studied analytically, as shown, for example, in Ref.s \cite{Hus:2013aa, Hus2013, Hus2014, Munao:2015vo}.

\subsection{The coarse-grained models for the mixture}

\begin{figure}
\centering
\includegraphics[width=0.8\columnwidth]{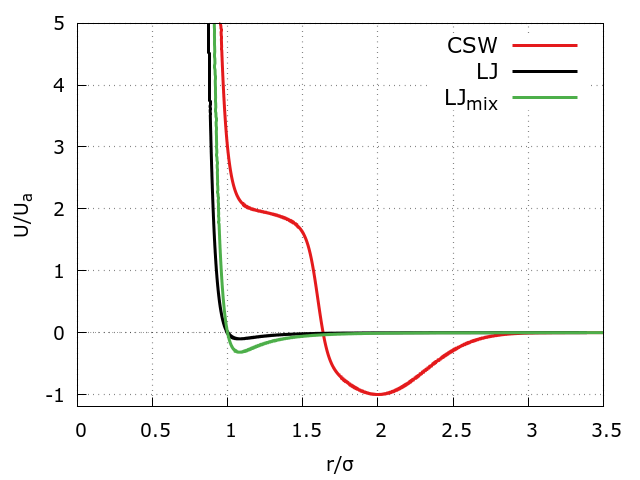}
\caption{{\bf Interactions potentials between the fluid beads.} 
The CSW potential (red line) has two characteristic length scales (the repulsive shoulder and the attractive well) as the hydrogen bond between polar groups in methanol and water.
The 24-6 Lennard-Jones (LJ, black line) and 24-6 LJ with Lorentz-Berthelot (LB) mixing rules (green line) are the interaction potentials for the methyl-methyl and methyl-hydroxyl interactions, respectively.}
\label{figpotentials}
\end{figure}

Following previous studies of water-methanol mixtures \cite{Marques:2020aa}, we represent a water molecule as a single (polar) CSW bead, while methanol as a dumbbell (two touching beads) made of an apolar 24-6 Lennard-Jones (LJ) bead for the CH$_3$ moiety and a polar CSW bead for the OH group. 
The CSW potential for the polar-polar (OH-[OH, H$_2$O], H$_2$O-H$_2$O) interactions is defined as (Fig. \ref{figpotentials}, red line)

\begin{equation} \label{eq:CSW}
    U_{\rm CSW}(r) \equiv \frac{U_R}{1+\exp \left( \frac{\Delta (r-R_R)}{a}\right)} - U_A \exp \left( - \frac{(r-R_A)^2}{2 \omega_A ^2 }\right) +U_A \left( \frac{a}{r} \right)^{24},
\end{equation}

with parameters

\begin{align*}
    \frac{U_R}{U_A}&=2, \, U_A=0.2 \, {\rm kcal/mol}, \\
    \frac{R_R}{a}&=1.6, \, a=1.77 \, \textup{\r{A}},\\
    \frac{R_A}{a}&=2, \, \left( \frac{\omega_A}{a} \right)^2=0.1, \\
    \Delta &= 15,
\end{align*}
where $a$ stands for the hard-core distance (the diameter of the particles); $R_R$ and $R_A$ are the repulsive radius and the distance of the attractive minimum, respectively; $U_R$ and $U_A$ are the energy of the repulsive shoulder and the attractive well, respectively; $\Delta$ controls the softness of the potential at $R_R$; and $\omega^2_A$ is the variance of the Gaussian centered in $R_A$ \cite{Fr07a, Vilaseca2010}. The values of the parameters for the CSW model are set in agreement with previous works \cite{Munao:2015vo, Hus:2014aa, Marques:2020aa, Leoni:2021aa}. Specifically, $U_A$ and $a$ are chosen to allow a comparison with atomistic water models \cite{Leoni:2021aa}, and $\Delta$ is as in Refs.~\cite{Munao:2015vo, Hus:2014aa} to benchmark our results against those for bulk.

The CH$_3$-CH$_3$ interaction (Fig. \ref{figpotentials}, black line) is described by the 24-6 LJ potential \cite{Munao:2015vo, Hus:2014aa, Marques:2020aa}
\begin{equation} \label{eq:LJ}
    U_{\rm LJ}(r)\equiv\frac{4}{3} 2^{2/3}\epsilon_{\rm LJ} \left[ \left(\frac{\sigma_{\rm LJ}}{r}\right)^{24} -\left( \frac{\sigma_{\rm LJ}}{r} \right)^6 \right],
\end{equation}
with
\begin{align*}
    \frac{\sigma_{\rm LJ}}{a}&=1.0, \hspace{0.2cm} \sigma_{\rm LJ}=1.77 \, \textup{\r{A}}, \\
    \frac{\epsilon_{\rm LJ}}{U_A}&=0.1, \hspace{0.2cm} \epsilon_{\rm LJ}=0.02 \, {\rm kcal/mol},
\end{align*}
where the values of $\sigma_{LJ}$ and $\epsilon_{LJ}$ are chosen such to compare with the CSW parameters.

The CH$_3$-[OH, H$_2$O] interaction is modeled with the 24-6 LJ potential employing the Lorentz-Berthelot mixing rules (Fig. \ref{figpotentials}, green line)
\begin{equation} \label{eq:LJ}
    U_{mix}(r)\equiv\frac{4}{3} 2^{2/3}\epsilon_{mix} \left[ \left(\frac{\sigma_{mix}}{r}\right)^{24} -\left( \frac{\sigma_{mix}}{r} \right)^6 \right],
\end{equation}
with
\begin{align*}
    \sigma_{mix}&\equiv\frac{1}{2}(\sigma_{\rm LJ}+a) = 1.77 \, \textup{\r{A}},\\
    \epsilon_{mix}&\equiv\sqrt{\epsilon_{\rm LJ}U_A} = 0.06 \, {\rm kcal/mol}.
\end{align*}

We consider three mixture compositions, 100\% CSW, 90\%-10\% CSW-methanol, and 75\%-25\% CSW-methanol. The first case allows us to establish a benchmark, while the other two can be compared against the bulk cases in Ref.~\cite{Marques:2020aa}. Regarding the relevance in actual cases, the 90\%-10\% composition can be compared to a mildly polluted water mixture. On the other hand, the more concentrated mixture with 75\%-25\% composition is possibly relevant in industry processes.

\subsection{The model for the graphene slit pore}

\begin{figure}
\centering
\includegraphics[width=0.8\columnwidth]{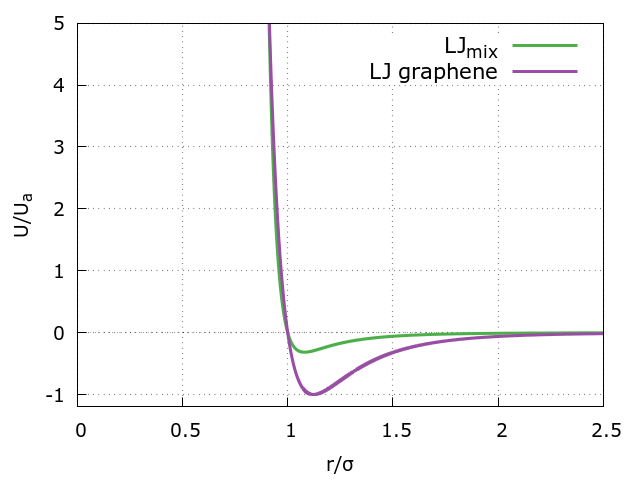}
\caption{{\bf Interactions potentials between the fluid beads and the graphene atoms.} 
The 12-6 LJ graphene-fluid interaction (violet line) is compared with the mixing potential in Fig.\ref{figpotentials} (green line). }
\label{figgrah_pot}
\end{figure}

Each graphene sheet is modeled as a honeycomb lattice in agreement with its atomic structure. 
Each graphene atom interacts with the fluid particles via the standard 12-6 LJ potential (Fig.\ref{figgrah_pot}, violet line)
\begin{equation}
    U_{\rm LJ}^{\rm graphene}(r)\equiv 4 \epsilon_g \left[ \left(\frac{\sigma_g}{r}\right)^{12}- \left( \frac{\sigma_g}{r} \right)^6 \right],
\end{equation}
with $\sigma_g=3.26\ \textup{\r{A}}$ and $\epsilon_g=0.1$ kcal/mol  \cite{Leoni:2021aa}, as established in literature \cite{https://doi.org/10.1002/jcc.20289}.
The positions of graphene atoms are kept fixed during the simulation.

\subsection{Graphene slit-pore geometry}

\begin{figure}
\centering
\includegraphics[width=0.6\columnwidth]{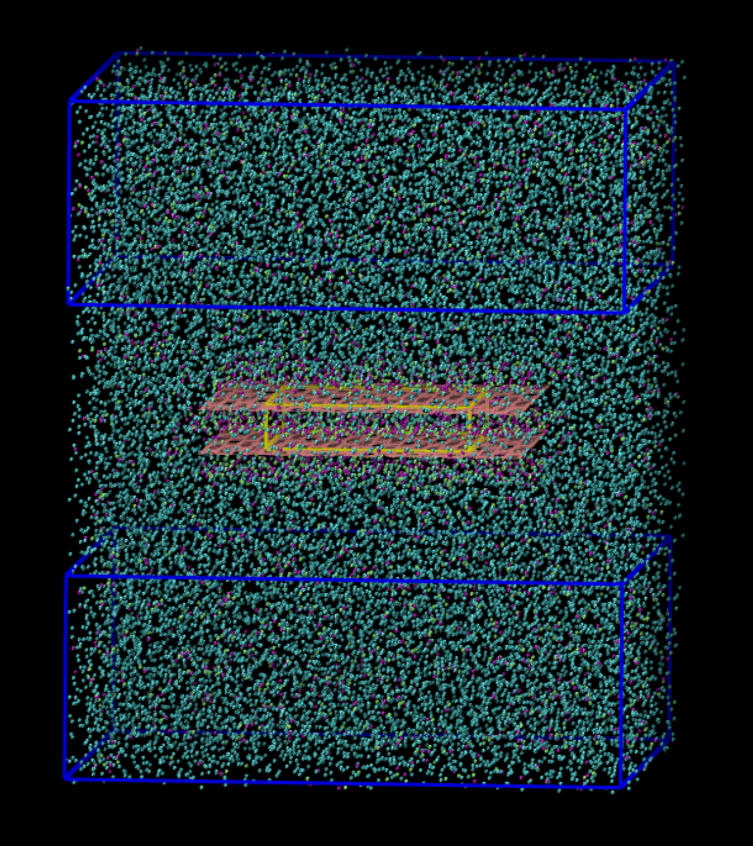}
\caption{{\bf Simulation snapshot for a 90\%-10\% CSW-methanol mixture}. The two navy blue boxes, connected by the periodic boundary conditions, correspond to the subregion with volume $V'$ in which we calculate the observables. 
Blue beads are CSW particles; methyl groups are purple; hydroxyl groups are green; pink lines represent the graphene lattice. The yellow box emphasizes the subvolume $V_s$ in which we compute the confined mixture composition.
}
\label{figsim}
\end{figure}

The graphene slit-pore is composed of two parallel sheets of sizes $l_{x, gr}=49\ \textup{\r{A}}$, $l_{y, gr}=51\ \textup{\r{A}}$ and width $l_{z, gr}=\delta$.
We vary $\delta$  from $6.5\,\textup{\r{A}}$ up to $13\,\textup{\r{A}}$.
The pore is included in a volume $V=L_x\, L_y\, L_z$,  with $L_x=L_y=84 $\AA\  and $L_z=98 $\AA\ and periodic boundary conditions (p.b.c.).

All the observables are calculated in a subvolume $V'\equiv L_{x}'\, L_{y}'\, L_{z}'$ sufficiently separated from the graphene walls to avoid direct interface effects,  with 
$L_{x}'=L_{y}'=84$ \AA, and $L_{z}'=44\ \textup{\r{A}}$ 
 (Fig. \ref{figsim}). With this choice of parameters, the distance of $V'$ from a graphene wall is always $>20$ \AA. 
 
To evaluate the changes of properties in $V'$, we compute the number density of the mixture and of both of its components as:
\begin{equation}
    \rho_\alpha\equiv  \frac{
    \exv{N_\alpha}
    }
    {V'},
\end{equation}
where $\exv{N_\alpha}$ is the ensemble average of the number of molecules $\alpha$, with $\alpha$ standing for CSW, methanol, or both (for which we use the symbol $\rho_{V'}$), inside the subregion $V'$. 
We also calculate the pressure of the system,
\begin{equation} \label{pressure}
    P_{V'}=\frac{\exv{N}k_BT}{V'} + \frac{1}{3V'}\exv{\sum_{i=1}^N\sum_{i>j}^N \mathbf{r}_{ij} \cdot \mathbf{f}_{ij}}.
\end{equation}
The first (kinetic) term comprises the ensemble average of all molecules $\exv{N}$, the Boltzmann constant $k_B$, the fixed temperature of the simulation $T$, and the volume of the subregion $V'$. The second term is the Virial, averaged over the ensemble of all the pairs of molecules $i$, $j$ in the volume $V'$, and $\mathbf{r}_{ij} \cdot \mathbf{f}_{ij}$ the scalar product between their distances and forces.
Finally, we compute the mixture composition in the subregion $V'$ outside the pore (Fig. \ref{figsim}).

For comparison, we also calculate the amount of methanol inside the pore as a function of the width $\delta$.
To minimize the edge effects of the walls, we consider only 
a reduced region of the slit-pore, i.e., a central subvolume $V_{s}=L_{x,s}\, L_{y,s}\,\delta$, where $L_{x,s}=L_{y,s}=30\ \textup{\r{A}}$.

\subsection{Molecular Dynamics}

We analyze the system by Molecular Dynamics (MD) simulations, where Newton's equations of motion of each degree of freedom are solved numerically in the canonical ensemble with a fixed total number $N=25,000$ of particles composing the mixture, in a total volume $V$, at fixed temperature $T=100$K controlled by the Nos\'e-Hoover thermostat, as implemented in the LAMMPS software \cite{LAMMPS}.
For the graphene slit-pore in the simulation box with p.b.c., we set the initial configuration by intertwining two hexagonal lattices (maximum packing) of CSW beads and methanol dimers and melting them during the initial equilibration procedure. Without the slit-pore, the number density would be $N/V=  0.036$ \AA$^{-3}$. 

We adopt the Leap-Frog integration algorithm \cite{Birdsall_2018}, a standard second-order and time-reversible integration method, with time-step $\delta t= 1$ fs.  We equilibrate the system for 0.1 ns to properly mix the two liquids and analyze the data collected every 100 fs for the next 0.1 ns.

\section{Summary and conclusions}

Water-methanol separation is relevant in several industrial applications, including methanol extraction for biofuels \cite{Ren_2000, Boysen_2004}. Still, traditional methods have limited efficiency and high economic costs  \cite{Liang_2014}. Therefore, exploring alternative approaches is technologically significant, scientifically challenging, and timely for the UN Sustainable Development Goals (SDGs)  ``Clean Water and Sanitation'' and ``Affordable and clean energy''. In particular, graphene sponges are novel nanomaterials that can effectively remove contaminants from water. They have several advantages over conventional methods, such as high surface area, tunable pore size, and multifunctional properties. Moreover, they can combine different water purification mechanisms, such as adsorption, catalytic, and electrocatalytic degradation of pollutants. This makes them promising candidates for various environmental engineering and water treatment applications \cite{Baptista_Pires_2021, Esplandiu:2023aa}.

Here, we investigate by Molecular Dynamics how an embedded graphene slit-pore modifies the properties of a water-methanol mixture. We find that the preferential interaction of the graphene with the hydrophobic moiety of the methanol induces an effective decrease of methanol concentration in the solution. 
On the other hand, 
the methanol accumulation inside the pore can be as high as 320\% of its nominal concentration under the condition we explore here, leading to a methanol depletion in the solution composition up to $\simeq 84$\% of its overall value.
Consequently, the methanol concentrations inside and outside the pore are anti-correlated with a more evident effect for higher methanol overall concentration (Fig.~\ref{figmet}).

We observe that the slit pore width $\delta$ has an appreciable effect on all the solution thermodynamic quantities of the mixture and its components outside the pore. In particular, their densities (Fig.~\ref{figbulk} a-c) and the mixture pressure (Fig.~\ref{figp_bulk}) decrease for increasing $\delta$. 
Therefore, these quantities can be tuned by changing the size of the embedded pore due to the size-dependent adsorption of methanol within the graphene-confined region.

Because we coarse-grain the mixture based on the CSW water-like liquid \cite{Fr07a} and the CSW-based dumbbell methanol model \cite{Hus:2014aa} to make the simulation efficient, we test if these results could be due to the approximation of our approach. Indeed,
both models have been tested in the literature \cite{oliveira2008, Munao:2015vo}, reproducing properties of the mixture  \cite{Marques:2020aa, Prslja:2019aa,Sentenac:1998aa,perry1999perry}.
Nevertheless, the lack of a directional interaction, leading to the formation of a hydrogen bond network, makes these models able to reproduce only part of the properties of the two components of the mixture, as discussed for the case of water, for example, in Ref.s~\cite{Vilaseca2011, Leoni:2021aa}.
We, therefore, test if the changes we find for the mixture density and pressure outside the pore as a function of the pore size are thermodynamically consistent. We find that they follow qualitatively the expected equation of state of the mixture, validating our results.
These results encourage further investigation to find the optimal parameters for graphene slit-pore and sponge applications in nano-filtering and purification of water-alcohol mixtures by physical mechanisms. In addition, further investigation will be necessary to understand if, for example, the sequestration of methanol within the graphene layers increases for an increasing number of layers.

\vspace{6pt} 


\authorcontributions{
Conceptualization, F.L, C.C.B., and G.F.; 
methodology, F.L, C.C.B., and G.F.; 
software, R.B.P.; 
validation, R.B.P., F.L, C.C.B., and G.F.; 
formal analysis, R.B.P.; 
investigation, R.B.P., F.L, C.C.B., and G.F.; 
resources, G.F.; 
data curation, R.B.P.; 
writing---original draft preparation, R.B.P., and G.F.; 
writing---review and editing, R.B.P., F.L, C.C.B., and G.F.; 
visualization, R.B.P., and G.F.; 
supervision, F.L, C.C.B., and G.F.; 
project administration, G.F.; 
funding acquisition, C.C.B., G.F. 
All authors have read and agreed to the published version of the manuscript.
}

\funding{
This research was funded by 
MCIN/AEI/ 10.13039/ 501100011033 and ``ERDF A way of making Europe" grant number PGC2018-099277-B-C22 and PID2021-124297NB-C31.
}

\dataavailability{
Data supporting reported results are available upon request.
} 

\acknowledgments{G.F. acknowledges the Visitor Program of the Max Planck Institute for The Physics of Complex Systems for supporting a six-month visit that started on November 2022. 
}

\conflictsofinterest{
The authors declare no conflict of interest.
The funders had no role in the design of the study; in the collection, analyses, or interpretation of data; in the writing of the manuscript; or in the decision to publish the~results.
} 


\abbreviations{Abbreviations}{
The following abbreviations are used in this manuscript:\\

\noindent 
\begin{tabular}{@{}ll}
MDPI & Multidisciplinary Digital Publishing Institute\\
CSW & Continuous shouldered well\\
MSD & Mean square displacement\\
CM & Center of mass\\
LJ & Lennard-Jones\\
MD & Molecular dynamics
\end{tabular}
}




\begin{adjustwidth}{-\extralength}{0cm}

\reftitle{References}


\bibliography{water-methanol}

\begin{thebibliography}{999}

\bibitem[Fetisov \em{et~al.}({\natexlab{a}})Fetisov, Gonopolsky, Zemenkova,
  Andrey, Davardoost, Mohammadi, and Riazi]{Fetisov_2023}
Fetisov, V.; Gonopolsky, A.M.; Zemenkova, M.Y.; Andrey, S.; Davardoost, H.;
  Mohammadi, A.H.; Riazi, M.
\newblock On the Integration of {CO}2 Capture Technologies for an Oil Refinery.
\newblock {\em 16},~865.
\newblock {\url{https://doi.org/10.3390/en16020865}}.

\bibitem[Fetisov \em{et~al.}({\natexlab{b}})Fetisov, Gonopolsky, Davardoost,
  Ghanbari, and Mohammadi]{Fetisov2023}
Fetisov, V.; Gonopolsky, A.M.; Davardoost, H.; Ghanbari, A.R.; Mohammadi, A.H.
\newblock Regulation and impact of {VOC} and {CO}$_2$ emissions on low-carbon
  energy systems resilient to climate change: A case study on an environmental
  issue in the oil and gas industry.
\newblock {\em Energy Science \& Engineering}, {\em 11},~1516--1535.
\newblock {\url{https://doi.org/10.1002/ese3.1383}}.

\bibitem[Soetens and Bopp(2015)]{Soetens:2015aa}
Soetens, J.C.; Bopp, P.A.
\newblock Water--Methanol Mixtures: Simulations of Mixing Properties over the
  Entire Range of Mole Fractions.
\newblock {\em The Journal of Physical Chemistry B} {\bf 2015}, {\em
  119},~8593--8599.
\newblock {\url{https://doi.org/10.1021/acs.jpcb.5b03344}}.

\bibitem[Ren \em{et~al.}()Ren, Zelenay, Thomas, Davey, and
  Gottesfeld]{Ren_2000}
Ren, X.; Zelenay, P.; Thomas, S.; Davey, J.; Gottesfeld, S.
\newblock Recent advances in direct methanol fuel cells at Los Alamos National
  Laboratory.
\newblock {\em 86},~111--116.
\newblock {\url{https://doi.org/10.1016/s0378-7753(99)00407-3}}.

\bibitem[Boysen \em{et~al.}()Boysen, Uda, Chisholm, and Haile]{Boysen_2004}
Boysen, D.A.; Uda, T.; Chisholm, C.R.I.; Haile, S.M.
\newblock High-Performance Solid Acid Fuel Cells Through Humidity
  Stabilization.
\newblock {\em 303},~68--70.
\newblock {\url{https://doi.org/10.1126/science.1090920}}.

\bibitem[Liu \em{et~al.}()Liu, Clemente, Hu, and Wei]{Liu_2007}
Liu, S.; Clemente, E.R.C.; Hu, T.; Wei, Y.
\newblock Study of spark ignition engine fueled with methanol/gasoline fuel
  blends.
\newblock {\em 27},~1904--1910.
\newblock {\url{https://doi.org/10.1016/j.applthermaleng.2006.12.024}}.

\bibitem[Yanju \em{et~al.}()Yanju, Shenghua, Hongsong, Rui, Jie, and
  Ying]{Yanju_2008}
Yanju, W.; Shenghua, L.; Hongsong, L.; Rui, Y.; Jie, L.; Ying, W.
\newblock Effects of Methanol/Gasoline Blends on a Spark Ignition Engine
  Performance and Emissions.
\newblock {\em 22},~1254--1259.
\newblock {\url{https://doi.org/10.1021/ef7003706}}.

\bibitem[Miganakallu \em{et~al.}(2020)Miganakallu, Yang, Rog{\'o}{\.z},
  Kapusta, Christensen, Barros, and Naber]{Miganakallu:2020aa}
Miganakallu, N.; Yang, Z.; Rog{\'o}{\.z}, R.; Kapusta, {\L}.J.; Christensen,
  C.; Barros, S.; Naber, J.
\newblock Effect of water - methanol blends on engine performance at borderline
  knock conditions in gasoline direct injection engines.
\newblock {\em Applied Energy} {\bf 2020}, {\em 264},~114750.
\newblock
  {\url{https://doi.org/https://doi.org/10.1016/j.apenergy.2020.114750}}.

\bibitem[Toledo-Camacho \em{et~al.}(2021)Toledo-Camacho, Rey, Maldonado,
  Llorca, Contreras, and Medina]{Toledo-Camacho:2021aa}
Toledo-Camacho, S.Y.; Rey, A.; Maldonado, M.I.; Llorca, J.; Contreras, S.;
  Medina, F.
\newblock Photocatalytic hydrogen production from water-methanol and -glycerol
  mixtures using Pd/TiO2(-WO3) catalysts and validation in a solar pilot plant.
\newblock {\em International Journal of Hydrogen Energy} {\bf 2021}, {\em
  46},~36152--36166.
\newblock
  {\url{https://doi.org/https://doi.org/10.1016/j.ijhydene.2021.08.141}}.

\bibitem[Miller and Carpenter(1964)]{Miller:1964aa}
Miller, G.A.; Carpenter, D.K.
\newblock Solid-Liquid Phase Diagram of the System Methanol-Water.
\newblock {\em Journal of Chemical \& Engineering Data} {\bf 1964}, {\em
  9},~371--373.
\newblock {\url{https://doi.org/10.1021/je60022a017}}.

\bibitem[Sun \em{et~al.}(2011)Sun, Wang, Tian, Liu, Ngai, and Tan]{Sun:2011aa}
Sun, M.; Wang, L.M.; Tian, Y.; Liu, R.; Ngai, K.L.; Tan, C.
\newblock Component Dynamics in Miscible Mixtures of Water and Methanol.
\newblock {\em The Journal of Physical Chemistry B} {\bf 2011}, {\em
  115},~8242--8248.
\newblock {\url{https://doi.org/10.1021/jp202893v}}.

\bibitem[Cortright \em{et~al.}(2002)Cortright, Davda, and
  Dumesic]{Cortright:2002aa}
Cortright, R.D.; Davda, R.R.; Dumesic, J.A.
\newblock Hydrogen from catalytic reforming of biomass-derived hydrocarbons in
  liquid water.
\newblock {\em Nature} {\bf 2002}, {\em 418},~964--967.
\newblock {\url{https://doi.org/10.1038/nature01009}}.

\bibitem[Masoumi and Dalai(2021)]{Masoumi:2021aa}
Masoumi, S.; Dalai, A.K.
\newblock Techno-economic and life cycle analysis of biofuel production via
  hydrothermal liquefaction of microalgae in a methanol-water system and
  catalytic hydrotreatment using hydrochar as a catalyst support.
\newblock {\em Biomass and Bioenergy} {\bf 2021}, {\em 151},~106168.
\newblock
  {\url{https://doi.org/https://doi.org/10.1016/j.biombioe.2021.106168}}.

\bibitem[Dalena \em{et~al.}(2018)Dalena, Senatore, Marino, Gordano, Basile, and
  Basile]{Dalena:2018aa}
Dalena, F.; Senatore, A.; Marino, A.; Gordano, A.; Basile, M.; Basile, A.,
  Chapter 1 - Methanol Production and Applications: An Overview.
\newblock In {\em Methanol}; Basile, A.; Dalena, F., Eds.; Elsevier,  2018; pp.
  3--28.
\newblock
  {\url{https://doi.org/https://doi.org/10.1016/B978-0-444-63903-5.00001-7}}.

\bibitem[Liang \em{et~al.}()Liang, Li, Luo, Xia, and Xu]{Liang_2014}
Liang, K.; Li, W.; Luo, H.; Xia, M.; Xu, C.
\newblock Energy-Efficient Extractive Distillation Process by Combining
  Preconcentration Column and Entrainer Recovery Column.
\newblock {\em 53},~7121--7131.
\newblock {\url{https://doi.org/10.1021/ie5002372}}.

\bibitem[Azamat(2019)]{Azamat:2019ua}
Azamat, J.
\newblock Selective separation of methanol-water mixture using functionalized
  boron nitride nanosheet membrane: a computer simulation study.
\newblock {\em Structural Chemistry} {\bf 2019}, {\em 30},~1451--1457.
\newblock {\url{https://doi.org/10.1007/s11224-019-01300-5}}.

\bibitem[Azamat \em{et~al.}(2021)Azamat, Ghasemi, Jahanbin~Sardroodi, and
  Jahanshahi]{Azamat:2021aa}
Azamat, J.; Ghasemi, F.; Jahanbin~Sardroodi, J.; Jahanshahi, D.
\newblock Molecular dynamics simulation of separation of water/methanol and
  water/ethanol mixture using boron nitride nanotubes.
\newblock {\em Journal of Molecular Liquids} {\bf 2021}, {\em 331},~115774.
\newblock {\url{https://doi.org/https://doi.org/10.1016/j.molliq.2021.115774}}.

\bibitem[Pr{\v s}lja \em{et~al.}(2019)Pr{\v s}lja, Lomba,
  G{\'o}mez-{\'A}lvarez, Urbi{\v c}, and Noya]{Prslja:2019aa}
Pr{\v s}lja, P.; Lomba, E.; G{\'o}mez-{\'A}lvarez, P.; Urbi{\v c}, T.; Noya,
  E.G.
\newblock Adsorption of water, methanol, and their mixtures in slit graphite
  pores.
\newblock {\em The Journal of Chemical Physics} {\bf 2019}, {\em 150},~024705.
\newblock {\url{https://doi.org/10.1063/1.5078603}}.

\bibitem[Mosaddeghi \em{et~al.}(2019)Mosaddeghi, Alavi, Kowsari, Najafi,
  Az'hari, and Afshar]{Mosaddeghi:2019aa}
Mosaddeghi, H.; Alavi, S.; Kowsari, M.H.; Najafi, B.; Az'hari, S.; Afshar, Y.
\newblock Molecular dynamics simulations of nano-confined methanol and
  methanol-water mixtures between infinite graphite plates: Structure and
  dynamics.
\newblock {\em The Journal of Chemical Physics} {\bf 2019}, {\em 150},~144510.
\newblock {\url{https://doi.org/10.1063/1.5088030}}.

\bibitem[Esplandiu \em{et~al.}(2023)Esplandiu, Bastus, Fraxedas, Ihmaz, Puntes,
  Radjenovic, Sep{\'u}lveda, Serr{\'a}, Su{\'a}rez-Garc{\'\i}a, and
  Franzese]{Esplandiu:2023aa}
Esplandiu, M.J.; Bastus, N.; Fraxedas, J.; Ihmaz, I.; Puntes, V.; Radjenovic,
  J.; Sep{\'u}lveda, B.; Serr{\'a}, A.; Su{\'a}rez-Garc{\'\i}a, S.; Franzese,
  G., Interfacial phenomena in nanotechnological applications for water
  remediation.
\newblock In {\em Reference Module in Chemistry, Molecular Sciences and
  Chemical Engineering}; Elsevier,  2023.
\newblock
  {\url{https://doi.org/https://doi.org/10.1016/B978-0-323-85669-0.00066-0}}.

\bibitem[Nair \em{et~al.}(2012)Nair, Wu, Jayaram, Grigorieva, and
  Geim]{Nair2012}
Nair, R.R.; Wu, H.A.; Jayaram, P.N.; Grigorieva, I.V.; Geim, A.K.
\newblock Unimpeded Permeation of Water Through Helium-Leak--Tight
  Graphene-Based Membranes.
\newblock {\em Science} {\bf 2012}, {\em 335},~442--444.

\bibitem[Mahmood \em{et~al.}()Mahmood, Bano, Kim, and Lee]{Mahmood2012}
Mahmood, A.; Bano, S.; Kim, S.G.; Lee, K.H.
\newblock Water--methanol separation characteristics of annealed SA/PVA complex
  membranes.
\newblock {\em 415-416},~360--367.
\newblock {\url{https://doi.org/https://doi.org/10.1016/j.memsci.2012.05.020}}.

\bibitem[Villegas \em{et~al.}()Villegas, {Castro Vidaurre}, and
  Gottifredi]{Villegas2015}
Villegas, M.; {Castro Vidaurre}, E.F.; Gottifredi, J.C.
\newblock Sorption and pervaporation of methanol/water mixtures with
  poly(3-hydroxybutyrate) membranes.
\newblock {\em 94},~254--265.
\newblock {\url{https://doi.org/https://doi.org/10.1016/j.cherd.2014.07.030}}.

\bibitem[Hung \em{et~al.}()Hung, Chang, Lecaros, Ji, An, Hu, Lee, and
  Lai]{Hung2017}
Hung, W.S.; Chang, S.M.; Lecaros, R.L.G.; Ji, Y.L.; An, Q.F.; Hu, C.C.; Lee,
  K.R.; Lai, J.Y.
\newblock Fabrication of hydrothermally reduced graphene oxide/chitosan
  composite membranes with a lamellar structure on methanol dehydration.
\newblock {\em 117},~112--119.
\newblock {\url{https://doi.org/https://doi.org/10.1016/j.carbon.2017.02.088}}.

\bibitem[Kachhadiya and Murthy()]{Kachhadiya2021}
Kachhadiya, D.D.; Murthy, Z.
\newblock Preparation and characterization of ZIF-8 and ZIF-67 incorporated
  poly(vinylidene fluoride) membranes for pervaporative separation of
  methanol/water mixtures.
\newblock {\em 22},~100591.
\newblock {\url{https://doi.org/https://doi.org/10.1016/j.mtchem.2021.100591}}.

\bibitem[Calero and Franzese(2020)]{calero2020}
Calero, C.; Franzese, G.
\newblock Water under extreme confinement in graphene: Oscillatory dynamics,
  structure, and hydration pressure explained as a function of the confinement
  width.
\newblock {\em Journal of Molecular Liquids} {\bf 2020}, {\em 317},~114027.
\newblock {\url{https://doi.org/https://doi.org/10.1016/j.molliq.2020.114027}}.

\bibitem[Leoni \em{et~al.}(2021)Leoni, Calero, and Franzese]{Leoni:2021aa}
Leoni, F.; Calero, C.; Franzese, G.
\newblock Nanoconfined Fluids: Uniqueness of Water Compared to Other Liquids.
\newblock {\em ACS Nano} {\bf 2021}, {\em 15},~19864--19876.
\newblock {\url{https://doi.org/10.1021/acsnano.1c07381}}.

\bibitem[Leoni and Franzese(2014)]{LF2014}
Leoni, F.; Franzese, G.
\newblock Structural behavior and dynamics of an anomalous fluid between
  attractive and repulsive walls: Templating, molding, and superdiffusion.
\newblock {\em The Journal of Chemical Physics} {\bf 2014}, {\em 141},~174501.
\newblock {\url{https://doi.org/http://dx.doi.org/10.1063/1.4899256}}.

\bibitem[Leoni and Franzese(2016)]{Leoni:2016aa}
Leoni, F.; Franzese, G.
\newblock Effects of confinement between attractive and repulsive walls on the
  thermodynamics of an anomalous fluid.
\newblock {\em Physical Review E} {\bf 2016}, {\em 94},~062604--.

\bibitem[Park \em{et~al.}()Park, Ni, C{\^o}t{\'e}, Choi, Huang, Uribe-Romo,
  Chae, O'Keeffe, and Yaghi]{Park2006}
Park, K.S.; Ni, Z.; C{\^o}t{\'e}, A.P.; Choi, J.Y.; Huang, R.; Uribe-Romo,
  F.J.; Chae, H.K.; O'Keeffe, M.; Yaghi, O.M.
\newblock Exceptional chemical and thermal stability of zeolitic imidazolate
  frameworks.
\newblock {\em 103},~10186--10191,
  \href{http://xxx.lanl.gov/abs/https://www.pnas.org/doi/pdf/10.1073/pnas.0602439103}{{\normalfont
  [https://www.pnas.org/doi/pdf/10.1073/pnas.0602439103]}}.
\newblock {\url{https://doi.org/10.1073/pnas.0602439103}}.

\bibitem[Franzese and Stanley(2010)]{life}
Franzese, G.; Stanley, H.E., Understanding the Unusual Properties of Water.
\newblock In {\em Water and Life: The Unique Properties of H$_2$O};
  Lynden-Bell, R.M.; Conway~Morris, S.; Barrow, J.D.; Finney, J.L.; Harper, C.,
  Eds.; CRC Press,  2010; chapter~7.

\bibitem[Bianco \em{et~al.}(2012)Bianco, Franzese, Ruberto, and
  Ancherbak]{fermi2012}
Bianco, V.; Franzese, G.; Ruberto, R.; Ancherbak, S.
\newblock Water and anomalous liquids.
\newblock In Proceedings of the Complex Materials in Physics and Biology;
  Mallamace, F.; Stanley, H.E., Eds. IOS Press,  2012, Vol. 176, {\em
  Proceedings of the International School of Physics "Enrico Fermi"}, pp.
  113--128.

\bibitem[Gallo \em{et~al.}(2021)Gallo, Bachler, Bove, B{\"o}hmer, Camisasca,
  Coronas, Corti, de~Almeida~Ribeiro, de~Koning, Franzese, Fuentes-Landete,
  Gainaru, Loerting, de~Oca, Poole, Rovere, Sciortino, Tonauer, and
  Appignanesi]{Gallo:2021wx}
Gallo, P.; Bachler, J.; Bove, L.E.; B{\"o}hmer, R.; Camisasca, G.; Coronas,
  L.E.; Corti, H.R.; de~Almeida~Ribeiro, I.; de~Koning, M.; Franzese, G.;
  et~al.
\newblock Advances in the study of supercooled water.
\newblock {\em The European Physical Journal E} {\bf 2021}, {\em 44},~143.
\newblock {\url{https://doi.org/10.1140/epje/s10189-021-00139-1}}.

\bibitem[Mart{\'\i} \em{et~al.}(2017)Mart{\'\i}, Calero, and Franzese]{MCF2017}
Mart{\'\i}, J.; Calero, C.; Franzese, G.
\newblock Structure and Dynamics of Water at Carbon-Based Interfaces.
\newblock {\em Entropy} {\bf 2017}, {\em 19},~135.
\newblock {\url{https://doi.org/10.3390/e19030135}}.

\bibitem[Corti \em{et~al.}(2021)Corti, Appignanesi, Barbosa, Bordin, Calero,
  Camisasca, Elola, Franzese, Gallo, Hassanali, Huang, Laria, Men{\'e}ndez,
  de~Oca, Longinotti, Rodriguez, Rovere, Scherlis, and Szleifer]{Corti:2021uy}
Corti, H.R.; Appignanesi, G.A.; Barbosa, M.C.; Bordin, J.R.; Calero, C.;
  Camisasca, G.; Elola, M.D.; Franzese, G.; Gallo, P.; Hassanali, A.;  et~al.
\newblock Structure and dynamics of nanoconfined water and aqueous solutions.
\newblock {\em The European Physical Journal E} {\bf 2021}, {\em 44},~136.
\newblock {\url{https://doi.org/10.1140/epje/s10189-021-00136-4}}.

\bibitem[Hu{\v s} and Urbic(2014)]{Hus:2014aa}
Hu{\v s}, M.; Urbic, T.
\newblock Existence of a liquid-liquid phase transition in methanol.
\newblock {\em Physical Review E} {\bf 2014}, {\em 90},~062306.

\bibitem[Baptista-Pires \em{et~al.}()Baptista-Pires, Norra, and
  Radjenovic]{Baptista_Pires_2021}
Baptista-Pires, L.; Norra, G.F.; Radjenovic, J.
\newblock Graphene-based sponges for electrochemical degradation of persistent
  organic contaminants.
\newblock {\em 203},~117492.
\newblock {\url{https://doi.org/10.1016/j.watres.2021.117492}}.

\bibitem[Sentenac \em{et~al.}(1998)Sentenac, Bur, Rauzy, and
  Berro]{Sentenac:1998aa}
Sentenac, P.; Bur, Y.; Rauzy, E.; Berro, C.
\newblock Density of Methanol + Water between 250 K and 440 K and up to 40 MPa
  and Vapor-Liquid Equilibria from 363 K to 440 K.
\newblock {\em Journal of Chemical \& Engineering Data} {\bf 1998}, {\em
  43},~592--600.
\newblock {\url{https://doi.org/10.1021/je970297p}}.

\bibitem[Perry \em{et~al.}(1999)Perry, Green, and Maloney]{perry1999perry}
Perry, R.; Green, D.; Maloney, J.
\newblock {\em Perry's Chemical Engineers' Handbook}; McGraw-Hill CD-ROM
  Handbooks, McGraw-Hill,  1999.

\bibitem[Marques \em{et~al.}(2020)Marques, Hernandes, Lomba, and
  Bordin]{Marques:2020aa}
Marques, M.S.; Hernandes, V.F.; Lomba, E.; Bordin, J.
\newblock Competing interactions near the liquid-liquid phase transition of
  core-softened water/methanol mixtures.
\newblock {\em Journal of Molecular Liquids} {\bf 2020}, {\em 320},~114420.
\newblock {\url{https://doi.org/https://doi.org/10.1016/j.molliq.2020.114420}}.

\bibitem[Vilaseca and Franzese(2011)]{Vilaseca2011}
Vilaseca, P.; Franzese, G.
\newblock Isotropic soft-core potentials with two characteristic length scales
  and anomalous behaviour.
\newblock {\em Journal of Non-Crystalline Solids} {\bf 2011}, {\em 357},~419 --
  426.
\newblock 6th International Discussion Meeting on Relaxation in Complex
  Systems, {\url{https://doi.org/DOI: 10.1016/j.jnoncrysol.2010.07.053}}.

\bibitem[Franzese(2007)]{Fr07a}
Franzese, G.
\newblock Differences between discontinuous and continuous soft-core attractive
  potentials: The appearance of density anomaly.
\newblock {\em Journal of Molecular Liquids} {\bf 2007}, {\em 136},~267--273.

\bibitem[Hu{\v s} \em{et~al.}(2014)Hu{\v s}, Muna{\`o}, and Urbic]{Hus2014}
Hu{\v s}, M.; Muna{\`o}, G.; Urbic, T.
\newblock Properties of a soft-core model of methanol: An integral equation
  theory and computer simulation study.
\newblock {\em The Journal of Chemical Physics} {\bf 2014}, {\em 141},~--.
\newblock {\url{https://doi.org/http://dx.doi.org/10.1063/1.4899316}}.

\bibitem[Hu{\v s} \em{et~al.}(2013)Hu{\v s}, Zalar, and Urbic]{Hus:2013aa}
Hu{\v s}, M.; Zalar, M.; Urbic, T.
\newblock Correctness of certain integral equation theories for core-softened
  fluids.
\newblock {\em The Journal of Chemical Physics} {\bf 2013}, {\em 138},~224508.
\newblock {\url{https://doi.org/10.1063/1.4809744}}.

\bibitem[Hus and Urbic(2013)]{Hus2013}
Hus, M.; Urbic, T.
\newblock Core-softened fluids as a model for water and the hydrophobic effect.
\newblock {\em The Journal of Chemical Physics} {\bf 2013}, {\em
  139},~114504--8.

\bibitem[Muna{\`o} and Urbic(2015)]{Munao:2015vo}
Muna{\`o}, G.; Urbic, T.
\newblock Structure and thermodynamics of core-softened models for alcohols.
\newblock {\em The Journal of Chemical Physics} {\bf 2015}, {\em 142},~214508.
\newblock {\url{https://doi.org/10.1063/1.4922164}}.

\bibitem[Vilaseca and Franzese(2010)]{Vilaseca2010}
Vilaseca, P.; Franzese, G.
\newblock Softness dependence of the anomalies for the continuous shouldered
  well potential.
\newblock {\em The Journal of Chemical Physics} {\bf 2010}, {\em 133},~084507.
\newblock {\url{https://doi.org/10.1063/1.3463424}}.

\bibitem[Phillips \em{et~al.}(2005)Phillips, Braun, Wang, Gumbart, Tajkhorshid,
  Villa, Chipot, Skeel, Kal{\'e}, and
  Schulten]{https://doi.org/10.1002/jcc.20289}
Phillips, J.C.; Braun, R.; Wang, W.; Gumbart, J.; Tajkhorshid, E.; Villa, E.;
  Chipot, C.; Skeel, R.D.; Kal{\'e}, L.; Schulten, K.
\newblock Scalable molecular dynamics with NAMD.
\newblock {\em Journal of Computational Chemistry} {\bf 2005}, {\em
  26},~1781--1802,
  \href{http://xxx.lanl.gov/abs/https://onlinelibrary.wiley.com/doi/pdf/10.1002/jcc.20289}{{\normalfont
  [https://onlinelibrary.wiley.com/doi/pdf/10.1002/jcc.20289]}}.
\newblock {\url{https://doi.org/https://doi.org/10.1002/jcc.20289}}.

\bibitem[Thompson \em{et~al.}(2022)Thompson, Aktulga, Berger, Bolintineanu,
  Brown, Crozier, in~'t Veld, Kohlmeyer, Moore, Nguyen, Shan, Stevens,
  Tranchida, Trott, and Plimpton]{LAMMPS}
Thompson, A.P.; Aktulga, H.M.; Berger, R.; Bolintineanu, D.S.; Brown, W.M.;
  Crozier, P.S.; in~'t Veld, P.J.; Kohlmeyer, A.; Moore, S.G.; Nguyen, T.D.;
  et~al.
\newblock {LAMMPS} - a flexible simulation tool for particle-based materials
  modeling at the atomic, meso, and continuum scales.
\newblock {\em Comp. Phys. Comm.} {\bf 2022}, {\em 271},~108171.
\newblock {\url{https://doi.org/10.1016/j.cpc.2021.108171}}.

\bibitem[Birdsall and Langdon()]{Birdsall_2018}
Birdsall, C.; Langdon, A.
\newblock {\em Plasma Physics via Computer Simulation}; {CRC} Press.
\newblock {\url{https://doi.org/10.1201/9781315275048}}.

\bibitem[de~Oliveira \em{et~al.}(2008)de~Oliveira, Franzese, Netz, and
  Barbosa]{oliveira2008}
de~Oliveira, A.B.; Franzese, G.; Netz, P.A.; Barbosa, M.C.
\newblock Waterlike hierarchy of anomalies in a continuous spherical shouldered
  potential.
\newblock {\em The Journal of Chemical Physics} {\bf 2008}, {\em 128},~064901.
\newblock {\url{https://doi.org/10.1063/1.2830706}}.

\end{thebibliography}

\PublishersNote{}

\clearpage




\end{adjustwidth}
\end{document}